\documentstyle[amssymb,aps,epsf]{revtex}

\begin{document}

\newcommand{\mafigura}[4]{
  \begin{figure}[hbtp]
    \begin{center}
      \epsfxsize=#1 \leavevmode \epsffile{#2}
    \end{center}
    \caption{#3}
    \label{#4}
  \end{figure} }

\author{Joaqu\'{\i}n Diaz-Alonso$^{1,3}$, Armando P\'erez$^{2}$ and 
Horacio D. Sivak$^{1}$}
\address{
$^{1}$ D.A.R.C., Observatoire de Paris-Meudon, UPR 176 CNRS
F-92195 Meudon, France.}
\address{$^{2}$ Departamento de F\'{\i}sica Te\'{o}rica, Universidad de Valencia
E-46100 Burjassot (Valencia) Spain.}
\address{$^{3}$ Departamento de F\'{\i}sica, Universidad de Oviedo
E-33007 Oviedo, Spain.}

\title{SCREENING EFFECTS IN RELATIVISTIC MODELS OF DENSE MATTER AT FINITE
TEMPERATURE}
\maketitle

\begin{abstract}
We investigate screening effects of the medium on the potential interaction
between two static 'charges' for different models of dense plasmas in the
one-boson exchange approximation. The potential can exhibit an oscillatory
behavior, which is related to the analytic structure of the corresponding boson
propagators in the complex $q$-plane. We have first revisited the one-pion
exchange in a nuclear medium. In addition to Friedel oscillations, which are
associated to branch cuts in the $q$-plane, there appears another oscillatory
component, which arises from a pole on the pion propagator. This pole is 
located
appart from the axes, giving rise to an oscillating Yukawa-like potential.
Therefore, we call this phenomenon 'Yukawa oscillations'. This phenomenon does
not appear in
the Debye component of the QED screened potential, even if the coupling
constant is artificially increased. We have also studied a model of QCD
quark-gluon plasma. In this case, the one-gluon propagator also shows this kind
of poles. At high densities and/or temperatures, where one expects perturbative
QCD to be valid, the pole shifts towards large momenta.
\end{abstract}

\section{\bf Introduction}

\begin{quote}
The natural framework for the analysis of high-density matter under conditions
where relativistic effects are essential is Quantum Field Theory \cite{BD65},
where dense plasmas are described (at fundamental or phenomenological level) as
assemblies of fermions (electrons, nucleons, quarks, etc...) interacting
through the exchange of bosons (photons, mesons, gluons, etc...). The dynamics
of these systems is defined by Lagrangian models whose solution (in general, in
a given approximation), allows for the calculation of macroscopic properties.
One interesting issue which can be obtained from the Lagrangian (using, for
example, the one-boson exchange approximation ''OBE'') is the description of
the two-body interaction in vacuum, in terms of a phenomenological
potential. Simple instances are the Coulomb and Yukawa 
potentials 
associated, respectively, to the exchange of massless and massive bosons.

A consequence of the collective behavior in a plasma is the appearance of
screening effects on the two-body interaction. In a classical
electromagnetic plasma, the Coulomb-like interaction between two static
charges in vacuum becomes, when modified by the medium, the exponentially
damped Debye interaction. However, in degenerate quantum electromagnetic
plasmas the in-medium potential becomes long-ranged and oscillatory, both in
the non-relativistic \cite{FR52} and relativistic \cite{SI85},\cite{De80}
cases. These so-called Friedel oscillations have observable consequences 
\cite{RO60}.

The phenomenological description of the nuclear interaction in vacuum through
the exchange of massive mesons \footnote{In the standard terminology, this
approach is known as 'Quantum Hadrodynamics' or QHD \cite{SW86}.} leads, in the
static OBE approximation, to an interparticle potential which is exponentially
damped with the distance \cite{M89}. The effects of the medium in a degenerate
nuclear plasma, as obtained in the same approximation \cite{GDP94}, or in a
direct calculation of the interaction energy \cite{DPS89}, introduce
qualitative modifications on the two-particle interaction potential, which
becomes long-ranged and oscillatory, as in the case of the degenerate
electromagnetic plasma. The analysis of the screened potential in both cases
allows for a separation into two components: a Friedel-like component, which is
damped as a power of the distance and dominates the long range, and an
exponentially damped Yukawa-like (or Debye-like in QED) component. We must 
stress, however, that there are qualitative differences between this Yukawa 
component
and the Debye screening of QED: the first one is, in fact, oscillatory below
some critical temperature, whereas not such oscillations are present in the
Debye component in the QED plasma. Friedel oscillations, which are present
in both cases, fade away for lower temperatures \cite{DPS89}. As we will
discuss later, these {\it Yukawa oscillations} are also present in other
physical systems such as a quark-gluon plasma, for which Kapusta et al.
\cite{KA88} have been evidenced the existence of Friedel oscillations. 

The analytic structure of the dressed propagators governs the behavior of
the screened potential \cite{DPS89}. From the mathematical point of view,
Friedel oscillations are associated to the Kohn singularity \cite{KO59} in
the matter part of the zero-temperature polarization tensor, which 
shows branch cuts starting
at a transferred momentum $q=\pm 2p_{f}$, where $p_{f}$ is the Fermi momentum
of fermions in the plasma. On the other hand, Yukawa
oscillations are associated to a pole of the propagator in the complex $q$
-plane which has a non-vanishing real part, in contrast to Debye screening,
which arises from a pole on the imaginary axis.

In this paper we analyze the two-particle interaction in dense plasmas, using
the above mentioned models. Here we are not interested in a precise
quantitative description of actual dense matter using more or less adapted
theoretical models. Our main concern is rather to evidence the above mentioned
characteristics of the screening, as a result from basic properties which 
seem to
be present in some relativistic models describing dense matter and, 
eventually, to discuss the limitations of such models.

The ground states of the plasma in the electromagnetic and in the quark-gluon
cases are described as  relativistic free Fermi gases in thermodynamical
equilibrium. For the QHD nuclear plasma, the ground state is described by the
relativistic Hartree approximation (RHA) \cite{CHIN77}, which defines the
thermodynamics at finite temperature and includes the contributions of 
vacuum fluctuations. The propagators of the exchanged bosons which are
necessary to obtain the screened potential are dressed by the medium. Their
expressions have been obtained in a self-consistent one-loop approximation,
where matter polarization contributions were calculated at finite
temperature and vacuum polarization contributions were included. Such 
approximation, for QHD,  is
the first order of a kinetic cluster expansion around the relativistic
Hartree ground state \cite{H78,DA85} and coincide at $T=0$ with the usual
one-loop approximation \cite{MS82}.

This paper is organized as follows.
In section 2 we introduce the essential equations which are necessary for
the analysis of the screening in the models. In section 3 we study the
analytic structure of the meson propagators and the oscillatory behavior of
the screened potentials. We conclude in section 4 with a discussion on the
eventual manifestations in actual systems of the phenomena found here.

\newpage
\end{quote}

\section{ Basic Equations}

\begin{quote}
\bigskip The static two-particle interaction potential in the 
medium can be
obtained from a linear response analysis, and can be put under the general
form \cite{KA88}: 

\begin{equation}
V(r)\ =\ \frac{Q_{1}Q_{2}}{4\pi ^{2}r}{\it Im}\int_{-\infty }^{\infty }dq%
\frac{qe^{iqr}}{q^{2}+F(\omega =0,q)}  \label{eq:(33)}
\end{equation}

\bigskip

Here, $r$ is the distance between the positions of the interacting
particles, $Q_{1}$ and $Q_{2}$ their corresponding ''charges'' and $%
k=(\omega ,\vec{q})$ the transferred four-momentum. A similar expression 
can be derived from the OBE approximation in the static ($\omega =0$ ) limit 
\cite{GDP94}. Then, $G(k)=\frac{-1}{-k^{2}+F(k)}$ corresponds to the boson
propagator, and $F(k)$ is related to the boson self-energy.

\bigskip

To be more explicit, we need to consider each case separately. As
representative of QHD, we will study the one-pion exchange. In this case,
one has to replace $Q_{1}Q_{2}\rightarrow -g_{\pi }^{2}$, with $g_{\pi}$ the
pion-nucleon coupling constant, and take into
account the pion mass. This means that the pion propagator $G_{\pi }(\omega
\ =\ 0,\vec{q})$ must be written as : 
\begin{equation}
G_{\pi }(\omega \ =\ 0,\vec{q})=-\frac{1}{q^{2}+m_{\pi }^{2}-g_{\sigma
\pi }m_{\sigma }<\sigma >-g_{\pi}^2 \Pi _{\pi }(q)}  \label{proppi}
\end{equation}

\bigskip

where $\Pi _{\pi }(k)\ $ is the pion polarization, which can be decomposed
into a matter and a renormalized vacuum contribution $\Pi _{\pi }(k)\ =\ \Pi
_{\pi }^{mat}(k)\ +\ \Pi _{\pi }^{vac}(k)$ and $m_{\pi }$ is the pion mass. 
We have included a pion-sigma coupling term in the Lagrangian, as in \cite
{DPS89}, with $g_{\sigma \pi }=3.77.$ Finally, $m_{\sigma }=550$ MeV represents
the $\sigma $-meson mass and $<\sigma >$ the self-consistent mean-field value of
this meson field. The renormalized vacuum contribution to the pion 
polarization is given by:

\begin{eqnarray}
\Pi _{\pi }^{vac}(k) &=&\ (2\pi )^{-2}\{-ak^{2}\log \left( \frac{1+a}{1-a}%
\right) +\ k^{2}\ -\ m_{\pi }^{2}\ +\ 2(m^{2}\ -\ M^{2})\   \nonumber \\
&&-\ 2(k^{2}\ -\ 2M^{2})\log (\frac{M}{m})-\ 4(m\ -\ M)^{2}(1\ +\
a_{0}^{-2})\ -\ (\frac{2}{a_{0}})\arctan (\frac{1}{a_{0}})  \nonumber \\
&&\times \left[ 2(m\ -\ M)^{2}(1\ +\ a_{0}^{-2})\ +\ k^{2}(1\ -\ \frac{2m^{2}%
}{m_{\pi }^{2}})\ -\ 2M^{2}\right] \}  \label{eq:(23)}
\end{eqnarray}

where $a\ =\ \sqrt{1\ -\ 4M^{2}/k^{2}}$ , $a_{0}\ =\ \sqrt{4m^{2}/m_{\pi
}^{2}\ -\ 1}$ and $M$ is the nucleon effective mass, calculated in the
Hartree approximation. In the next section we give the expression for the
matter contribution. 

From the expression (\ref{eq:(33)}) we see that the analytic structure of
the propagators on the $\omega \ =\ 0$ axis becomes the essential magnitude
which governs the behavior of the screening. In the QHD case there are poles
of the propagator on this axis associated to the ''tachyonic'' branches,
which come from the vacuum polarization terms \cite{DP91}. Such poles arise
at large momentum transfer, where the internal structure of the nucleon
invalidates the point-particle approach of the model and thus, the effect of
this pole on the calculated in-medium potential is spurious. We can try to
account phenomenologically for the nucleon structure by the introduction of
monopolar form factors $ff(k)$ on the vertex of the loop expansion and the
OBE diagrams. To do this we have to replace, in the previous equations

\begin{eqnarray}
g_{\pi }^2 &\rightarrow &g_{\pi }^2 ff(k)  \nonumber \\
ff(k) &=&(\Lambda _{\pi }^{2}\ -\ \mu _{\pi }^{2})/(\Lambda _{\pi }^{2}\ -\
k^{2})  \label{eq:(34)}
\end{eqnarray}

where $\Lambda _{\pi }$ is a cut-off parameter. When the value of this
parameter is properly chosen, the spurious ''tachyonic'' branches (and the
associated pole of the propagator on the $\omega =0$ axis) disappear. Of
course, one
could also eliminate the poles associated to these branches by simply
discarding the renormalized vacuum polarization contribution. However, this
contribution is essential in obtaining the physical propagation modes. If
neglected, new unphysical space-like modes (which introduce new poles in the
propagators at small $q$) are present and can not be eliminated by form
factors \cite{DP91}. The standard expression of the form factors Eq. (\ref
{eq:(34)}) fits well the known data in vacuum \cite{M89}. At finite density
and temperature the form factor is expected to be modified by the medium. In
a simple bag-model picture the nucleons ''swell'' as density increases (EMC
effect), due to a reduction of the bag pressure. Higher-order calculations
of the vertex function \cite{AS92}, where the surrounding meson cloud
defines the size of the nucleon, leads to a similar behavior. Here we shall
keep the form factor unchanged in the medium according to our level of
approximation, but it must be emphasized that, in an improved calculation,
the above mentioned effect will cause a stronger suppression of
high-momentum exchanges at larger densities. Nevertheless, these
modifications would affect the short-range part of the interaction, and will
not modify our main conclusions, which concern the long-range behavior of
the screened potential.

We will also analyze the one-gluon exchange screened potential on a
color-singlet state of a quark-antiquark pair. For this analysis, we have
considered a $SU(N)$ quark-gluon plasma with $N_{f}$ quark flavors. One has
to replace 
\begin{equation}
Q_{1}Q_{2}\rightarrow -\frac{(N^{2}-1)g^{2}}{2N}  \label{qqcd}
\end{equation}
in this case in Eq. (\ref{eq:(33)}), where $g$ will be the coupling constant. In 
order
to simplify our analysis , we have restricted ourselves to a plasma with only
$u$ and $d$ massless quarks with the same chemical potential. The denominator
in Eq. (\ref {eq:(33)}) is then related to the one-loop gluon polarization in the
static limit. The corresponding formulae can be found, for example, in
\cite{KA89},\cite{KA85} and \cite{TO85}. We will write explicit expressions
for zero temperature in the next section.
\end{quote}

\section{\bf \ Analysis of the in-medium interactions.}

\begin{quote}
As discussed in the previous section, in the case of the pion-nucleon
plasma, one has to evaluate Eq.(\ref{eq:(33)}) with the 
quasi-pion propagator given by Eq.(\ref{proppi}) and the introduction of the
form factors given by the replacement (\ref{eq:(34)}). The integration can
be transformed to a contour integral in the complex $q$-plane by analytical
continuation of $\Pi (q)\equiv $ $\Pi (\omega =0,q)$. Let us first discuss
the situation at zero temperature. We will summarize the main results found
in \cite{DPS89}. The matter contribution can be written as: 
\begin{eqnarray}
\Pi _{\pi }^{mat}(q) &=&-(2\pi )^{-2}\left[ 4\varepsilon _{f}p_{f}-\left(
4M^{2}+2q^{2}\right) \log \left( \frac{\varepsilon _{f}+p_{f}}{M}\right)
+2\varepsilon _{f}q\log \left( \frac{q-2p_{f}}{q+2p_{f}}\right) \right.  
\nonumber \\
&&\left. +aq^{2}\log \left( \frac{ap_{f}+\varepsilon _{f}}{%
ap_{f}-\varepsilon _{f}}\right) \right]   \label{polmat}
\end{eqnarray}
Now $a\ =\ \sqrt{1\ +\ 4M^{2}/q^{2}}$ (we have corrected two misprints
appearing in Eq.(4.8) of reference \cite{DPS89}). One has on
the upper-half plane the analytical structure of the pion propagator shown 
in Fig. 1a)\footnote{
Aside the branch cuts at $q=\pm 2p_{f}$ , there are also branch cuts at $%
q=\pm 2iM$ (not shown in the figure). They correspond to the thereshold of
particle-antiparticle production. However, they will contribute only for the
short-distance range ($r\lesssim 1fm$). Here, we will neglect 
this contribution.}%
, where we have also shown the contour path to be used in evaluating the
integral Eq. (\ref{eq:(33)}) . The branch cuts starting at $q=\pm 2p_{f}$
appear from the logarithms in Eq.(\ref{polmat}) and have their origin on the
sharp form of the Fermi surface in momentum space at $T=0$. They give rise
to Friedel oscillations. For the nuclear plasma, they could manifest as an
instability of the uniform configuration in the Hartree approximation, thus
allowing for the possibility of new phases in nuclear matter\cite{Pri90i},%
\cite{Pri90ii}. Alternatively to the use of contour integrals, one can
integrate by parts Eq. (\ref{eq:(33)}) , following a systematic method given
by Lighthill \cite{LI64} ; then one finds that $\Pi _{\pi }^{mat}(q)$ has a
pole on the second derivative : the Kohn singularity \cite{KO59}. In
addition to this, one has the pole contributions at the points $q=\pm
q_{r}+iq_{i}$ shown in Fig. 1a) as the small crosses with a circular contour
around them, which we will refer to as the {\it Yukawa pole}. Then the
potential $V(r)$ , for large $r$ will consist on two terms : 
\begin{equation}
V(r\rightarrow \infty )\sim V_{F}(r)\ +\ V_{Y}(r)
\end{equation}

The first component $V_{F}(r)$ is the Friedel component, which is
oscillatory and decays as a power-law of $r$ for large distances, while $%
V_{Y}(r)$ is the oscillatory exponentially-damped term (Yukawa term), which
has the form :

\begin{equation}
V_{Y}(r)\ =\ A\frac{1}{r}\sin (q_{r}r)e^{-q_{i}r}
\end{equation}

where the amplitude $A$, the ''effective'' pion mass $q_{i}$ and the
''frequency'' $q_{r}$ depend on the plasma density. At low densities, the
frequency $q_{r}$ becomes imaginary and the oscillation disappears. When
density vanishes this component reduces to the usual Yukawa potential
\cite{DPS89}.

\bigskip We will now discuss the situation for finite temperature. At
non-zero temperature $T$, the Fermi surface is spread over a thickness $T$
and the Kohn singularity disappears. However, at small temperatures the
second derivative in the polarization $\Pi _{\pi }(q,T)$ keeps still an
important jump around $q\ =\ \pm 2p_{f}$. This can be seen in Fig. 2, where we
have plotted the second derivative of the polarization as a function of $q/m$%
, for various temperatures at saturation density (which corresponds to $%
p_{f}\simeq 0.3m$ ). We see the Kohn singularity at $T\ =\ 0$ and the
evolution of the jump around $q\ =\ 2p_{f}$ as temperature increases. For
temperatures around 20 MeV the propagator is a smooth function of $q.$

At low but finite temperatures ($(\varepsilon _{f}\ -\ M)/T\gg 1$), with 
$\varepsilon _{f}$  the nucleon Fermi energy, the Friedel potential becomes

\begin{equation}
V_{F}(r,T)\approx B(Tp_{f}^{2}/r^{2})/\nu ^{2}(2p_{f},T)\cos (2p_{f}r)/\sinh
(\frac{2\pi rT\varepsilon _{f}}{p_{f}})  \label{VF}
\end{equation}
where

\begin{equation}
\nu (q,T)\ =\ (q^{2}\ +\ \Lambda _{\pi }^{2})(q^{2}\ +\ m_{\pi }^{2})\ -\
g_{\pi }^{2}(\Lambda _{\pi }^{2}\ -\ m_{\pi }^{2})\Pi _{\pi }(q,T)
\label{eq:(37)}
\end{equation}
and $B$ is a constant. In the limit $T\ =\ 0$ , Eq.(\ref{VF}) behaves as 
\begin{equation}
V_{F}(r)\ \approx \ cos(2P_{f}r)/r^{3}\   \label{VFcos}
\end{equation}

As temperature increases, the Friedel component is exponentially reduced
with $T$ (see Eq.(\ref{VF})).
In the present pion-nucleon model this component fades away, at saturation
density, as temperature approaches $T\sim 20MeV.$ 

The oscillations associated to the Yukawa component are present up to $T\sim
60MeV$. We have shown this behavior in the next figures, where we plot the
real part $q_{r}$ (Fig. 3) and the imaginary part $q_{i}$ (Fig. 4) of the
Yukawa pole, both in units of the free nucleon mass $m$ as a function of the
temperature, for three values of the density, defined by a given $p_{f}$ at 
$T=0$ : $p_{f}=0.3$, corresponding to saturation density (solid line), 
$p_{f}=0.35$ (dotted line) and $p_{f}=0.4$ (dashed line). As it is readily
seen, the pole shifts towards the imaginary axis as temperature increases, and
remains as a pure imaginary pole for temperatures above some critical value
$T_{c}$, which depends on the density. As a consequence, Yukawa oscillations
are present only if $T<T_{c}$. This transition can also be presented, from a
formal point of view, as $ q_{r}$ 'becoming imaginary'(and negative), thus
being effectively substracted from the value of $q_{i}$ and producing the kink
observed on Fig. 4. 

Now the question is wether the Yukawa pole found in the pion propagator is a
peculiarity of this model or not. We have performed a similar study for the
screened photon propagator in QED and the screened gluon propagator in QCD.
In the first case, we found a pole on the imaginary axis for all the values
of the density and temperature we considered. Of course, this imaginary pole
corresponds to the well-known Debye screening phenomenon. An important
difference with QHD is the strength of the coupling constant : in QED one
has $\alpha =\frac{1}{137}$, whereas for the pion-nucleon model it is much
larger ($\frac{g_{\pi }^{2}}{4\pi }\sim 14 $). We have artificially increased
the value of the QED coupling constant, by taking values $\alpha \sim 10$ in
order to test if the Yukawa pole arises in QHD as a consequence of 
the large value of the coupling constant. However, even in this case, we have
not found such pole in the electromagnetic case. Therefore we conclude that
the strength of the coupling constant is not the main reason for the
appearance of Yukawa oscillations, and they must be instead related to the
nature of the model. 

\bigskip

The situation is completely different in QCD. Let us consider the gluon
propagator at the one-loop approximation, for the model described in the
previous section. The denominator in Eq.(\ref{eq:(33)})
can be written as : 

\begin{equation}
D(q)\equiv q^{2}+F(q)=q^{2}+F^{mat}(q)+F^{vac}(q)  \label{propgluo}
\end{equation}

where $F^{mat}(q)$ and $F^{vac}(q)$ stand for the matter and vacuum
contributions, respectively. We follow the formulae given in \cite{KA88} to
write these contributions (calculations in this reference are made using the
temporal axial gauge). 
At zero temperature, the matter term is given by 

\begin{equation}
F^{mat}(q)=\frac{g^{2}}{48\pi ^{2}}N_{f}\left[ 16p_{f}^{2}+\frac{2p_{f}}{q}%
\left( 4p_{f}^{2}-3q^{2}\right) \log \left( \frac{2p_{f}+q}{2p_{f}-q}\right)
-2q^{2}\log \left( \frac{q^{2}-4p_{f}^{2}}{q^{2}}\right) \right]
\label{matgluo}
\end{equation}

and the remaining terms can be combined to give 
\begin{equation}
q^{2}+F^{vac}(q)=\frac{g^{2}}{48\pi ^{2}}(11N-2N_{f})q^{2}\log \left( \frac{%
q^{2}}{\Lambda _{QCD}^{2}}\right)  \label{vacgluo}
\end{equation}

Using Eqs. (\ref{eq:(33)}),(\ref{qqcd}),(\ref{propgluo}) and factoring out $%
\frac{g^{2}}{48\pi ^{2}}$ by defining 
\begin{equation}
D(q)\equiv \frac{g^{2}}{48\pi ^{2}}f(q)  \label{factorg}
\end{equation}
one finds: 
\begin{equation}
V(r)\ =-\ \frac{6\left( N^{2}-1\right) }{Nr}{\it Im}\int_{-\infty }^{\infty
}dq\frac{qe^{iqr}}{f(q)}  \label{vqcd}
\end{equation}

Thus, the potential $V(r)$ is independent of $g^{2}$. At zero
temperature, the analytic structure of the gluon propagator is shown in Fig.
1b). Aside the branch cuts at $q=\pm 2p_{f}$ , there is also a branch cut on
the imaginary axis.

At zero density ($p_{f}=0$), $F^{mat}(q)$ vanishes, and one has 
\begin{equation}
f(q)=(11N-2N_{f})q^{2}\log \left( \frac{q^{2}}{\Lambda _{QCD}^{2}}\right) 
\label{propvac}
\end{equation}
Therefore, the gluon propagator has poles on the real axis (at $q=\pm
\Lambda _{QCD}$ ), and the long-distance behavior of $V(r)$ is dominated by
these poles. We have analyzed the behavior of this pole as $p_{f}$
increases, for a chosen value of $\Lambda _{QCD}=200$ MeV. First, the
position of this pole shifts from the value of $\Lambda _{QCD}$, and a
second pole at smaller $q$ appears. However, for $p_{f}\gtrsim 120$ MeV both
of them disappear and $1/D(q)$ will not have poles on the real axis. Since
the deconfinement transition is thought to take place (at $T=0$) at a
density such that $p_{f}\gtrsim 300$ MeV, we can assume that no poles on
the real axis will be present in the quark-gluon plasma phase. Instead, a
pole with both a real and an imaginary part appears at $q_{0}=q_{r}+iq_{i}$,
as shown in Fig. 1b). This means that we have a situation similar to that
encountered in the one-pion exchange, and we could have Yukawa-like
oscillations. However, some remarks are in order. At zero temperature and $%
p_{f}\thickapprox 300$ MeV we found $|q_{0}|\thickapprox 200$ MeV, which is
the value of $\Lambda _{QCD}$. This seems to invalidate our one-loop
calculation, since we expect non-perturbative effects to be very large under
these circumstances. Nevertheless, $p_{f}$ introduces a new energy-scale
into the problem (apart from $\Lambda _{QCD}$) and therefore the situation
may be different for large values of $p_{f}$. In order to investigate this
possibility, we have calculated the evolution of $q_{0}$ for increasing
densities. The result can be seen in Fig. 5, where we have represented both
the real part (solid line) and the imaginary part (dashed line) of $q_{0}$ as
a function of $p_{f}$ (all units are in GeV). As one can see, for large
values of $p_{f}$ they become larger than $\Lambda _{QCD}$, making the
perturbative expansion hopefully more reliable.

A similar situation appears if one keeps the density constant and increases
the temperature. In this case, one has to recalculate Eq. (\ref{matgluo}) to
take into account the temperature effects. The appropriate formulae can be
found in \cite{KA89},\cite{KA85},\cite{TO85}. We address the reader to these
references and will not repeat here these formulae.

We have investigated the evolution of the Yukawa pole with temperature, for
a fixed quark particle (minus antiparticle) density which corresponds to $%
p_{f}=300$ MeV at $T=0$. This is plotted in Fig. 6, where we followed
similar conventions to the above figure, but now in the horizontal axis one
has the temperature (in GeV). As we found before, large values of $T$ makes
the pole appear at large momenta. We made similar calculations starting from
larger values of $p_{f}$ and found no qualitative differences with the
previous case. One expects the perturbative expansion to work better under
these conditions. In fact, the asymptotic freedom property of QCD should
manifest in this regime.

Let us now concentrate on the physical consequences of the analytic
structure of the gluon propagator at $T=0$ for the large-distances potential 
$V(r)$. By applying the residue theorem, the pole (Yukawa) contribution can
be written as :

\begin{equation}
V_{Y}(r)=-\frac{24\pi (N^{2}-1)}{N\left( a^{2}+b^{2}\right) }\frac{\exp
(-q_{i}r)}{r}\left[ a\cos (q_{r}r)+b\sin (q_{r}r)\right]   \label{vyuk}
\end{equation}

where 
\begin{equation}
\frac{f^{^{\prime }}(q_{0})}{q_{0}}\equiv a+ib
\end{equation}

has to be calculated at the position of the pole . One has also to include
the contribution to $V(r)$ coming from the branch cut running up the
imaginary axis, which we will write as $V_{im}(r)$. At zero temperature and
large distances, this contribution can be easily approximated (proceeding as
in \cite{KA88}) :

\begin{equation}
V_{im}(r\rightarrow \infty )\simeq \frac{(11N-2N_{f})g^{4}}{32N\pi ^{3}}%
\frac{(N^{2}-1)}{m_{el}^{4}r^{5}}=\frac{\pi \left( 11N-2N_{f}\right) }{8N}%
\frac{\left( N^{2}-1\right) }{N_{f}^{2}p_{f}^{4}r^{5}}  \label{contribim}
\end{equation}

with the electric mass defined as

\begin{equation}
m_{el}^{2}=D(q\rightarrow 0)=\frac{g^{2}N_{f}p_{f}^{2}}{2\pi ^{2}}
\label{mel}
\end{equation}

The contribution from the branch cuts at $q= \pm 2 p_f$ (Friedel component)
is similar to Eq. (\ref{VFcos}) (see \cite{KA88}).  
In Fig. 7 we have plotted the 
different components of $V(r)$ (in MeV)
mentioned above, at zero temperature and $p_{f}=300$ MeV, as a function of
the distance (in fm). The solid line gives the result of the numerical
integration of Eq. (\ref{vqcd}) : we will call this the {\it exact}
potential. The dotted line shows the Yukawa component, as given by Eq. (\ref
{vyuk}) . Also shown is the contribution from Eq. (\ref{contribim}) (long
dashes) and the Friedel component, as given in Ref. \cite{KA88} (short dashes).
Finally, the dotted-dashed line is obtained by the sum of the Friedel, Yukawa
and Eq. (\ref{contribim}). As can be seen from this figure, the Yukawa
component alone gives a reasonable approximation to $V(r)$ for this distance
range. Also, it is clear that Eq. (\ref{contribim}) overestimates the imaginary
branch cut contribution for moderate distances. The Friedel component here is
much smaller than the others. 

\bigskip

The situation changes for larger distances. We have represented in Fig. 8
the same pieces of the potential, but for a distance range $r:5-10$ fm. Now
the contribution from $V_{im}(r)$ makes the approximate potential closer to
the exact curve. This becomes apparent when we consider even a larger $r$ ,
as seen in Fig. 9. For this range of
distances, the Yukawa component tends to disappear, while Friedel oscillations
are still important, so that the potential can be approximated by the
Friedel and $V_{im}(r)$ contributions. Anyway, for distances $r \gtrsim 5$
fm, $V(r)$ becomes too small. For the distance range where the potential is
of the order of a few MeV, it is reasonably approximated by the Yukawa
component alone.

\end{quote}

\section{Conclusions}

In this work, we have considered plasma screening effects on the potential
interaction for two static 'charges' in different models of dense matter.
First, we studied
the one-pion exchange in nuclear matter, as representative of QHD (Quantum
Hadrodynamics). In addition to Friedel oscillations, which arise due to
logarithmic branch cuts on the complex q-plane of the pion propagator, we
find that this propagator shows a pole on this plane at zero temperature and
densities above saturation. This pole gives rise to additional oscillations
on the nucleon-nucleon potential, which we refer to as 'Yukawa
oscillations'. In contrast to Friedel oscillations, which are damped as some
power-law of the distance, Yukawa oscillations are exponentially damped with
distance. Another important difference is the temperature behavior. While
Friedel oscillations are strongly suppressed for temperatures of a few MeV,
Yukawa oscillations are present up to $T\sim 60$ MeV when the density is
equal to the saturation density. There is a critical temperature $T_{c}$ for
each density, below which Yukawa oscillations are present.

In order to investigate wether this phenomena are a peculiarity of this
model or not, we have made further research. One feature of the pion
exchange is the large value of the coupling constant. We considered the
one-photon propagator in a QED plasma with an artificially increased value
of the electromagnetic coupling constant. However, we found no Yukawa poles
even if the coupling constant is taken to be of the order of unity or more. 
Thus,
a large value of the coupling constant alone can not explain the origin of
such pole.

Another important feature of QHD is that it shows no asymptotic freedom.
This feature originates some pathologies in the meson sector, such as the
appearance of tachyonich branches \cite{Pe87},\cite{We90}. Therefore, one
might suspect that Yukawa poles are also a pathological behavior of
non asymptotically-free field-nuclear models. In order to investigate this
aspect, we have studied the singlet quark-antiquark potential in a QCD
quark-gluon plasma. In the case we studied, Yukawa poles appear, at zero
temperature, when the quark Fermi momentum reaches $p_{f}\gtrsim 120$ MeV
and remain for larger densities. When the temperature is increased they
survive, at least up to $T=2$ GeV, which is the maximum temperature we
reached in our calculations. This situation is in contrast with the one-pion
model, where the pole disappears at some critical temperature. It is
important to notice that, although the values of $q$ at the pole are close
to $\Lambda _{QCD}$ for low densities and temperatures, they become large
(as compared to $\Lambda _{QCD}$) if the density and/or temperature
increases. Since one expects to obtain asymptotic freedom in this limit, a
perturbative treatment like the one we have considered has more chances to
be valid in these circumstances. In other words, Yukawa poles in a nuclear
medium do not seem to be associated to non-asymptotically free theories and
might correspond to some underlying physical phenomenon. As in the case of
Friedel oscillations, this could manifest in the appearance of new phases in
dense matter. Of course, they could also be a consequence of the
approximations made in our models (namely the one-loop approximation) and
might be not present in more realistic calculations. This will be the
subject of a future work.

{\bf Acknowledgments}

We are grateful to Drs. V. Vento and A. Nieto for fruitful discussions. This
work has been partially supported by Spanish DGICYT Grant PB94-0973 and
CICYT AEN96-1718.

\mafigura{6cm}{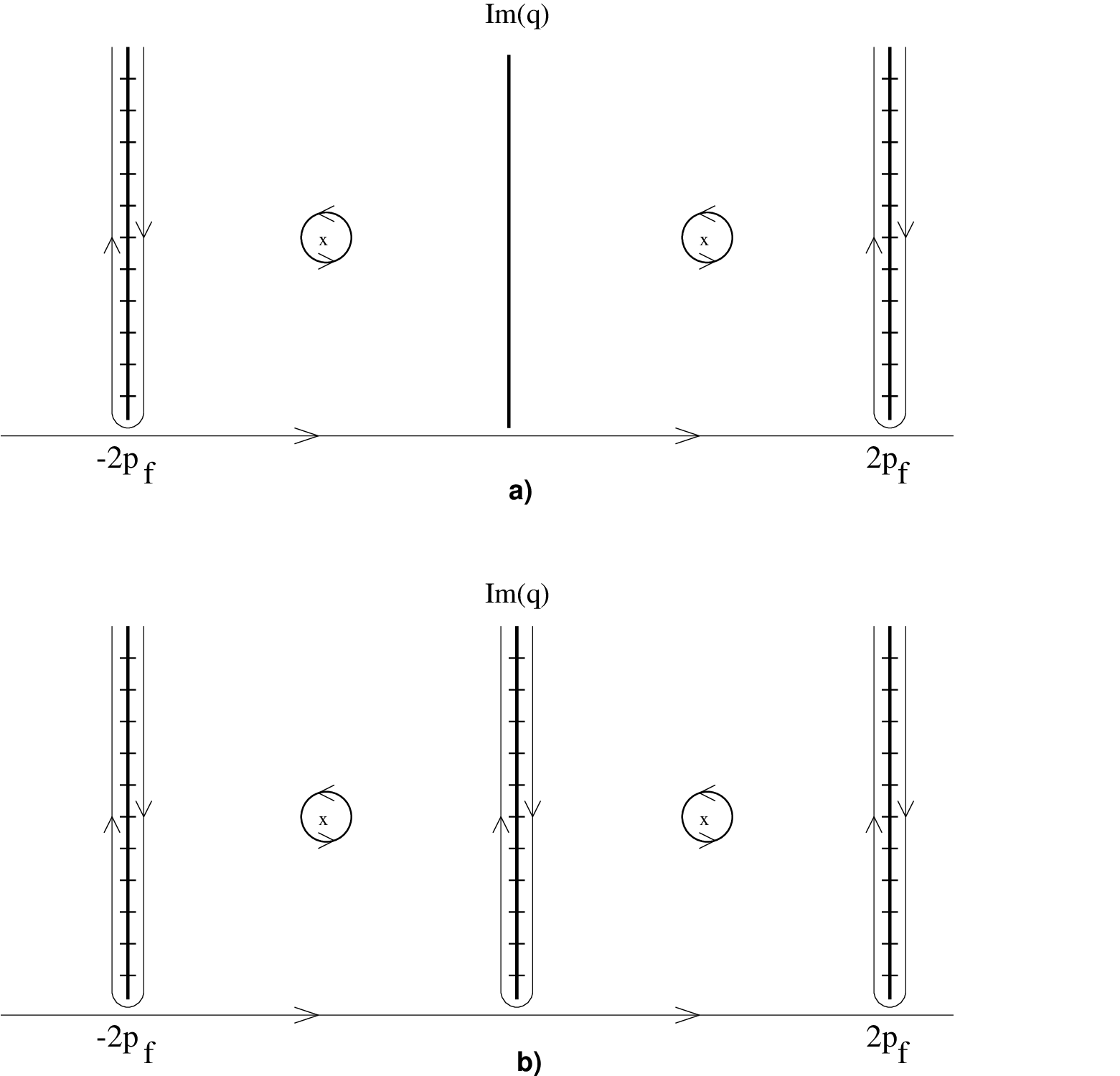}{Analytically-continued structure of the pion
propagator (upper part) and gluon propagator (lower part) at zero 
temperature. Represented schematically are the 
contour paths used for integration. The Yukawa pole is indicated as an 
encircled cross.}{Fig. 1} 

\mafigura{6cm}{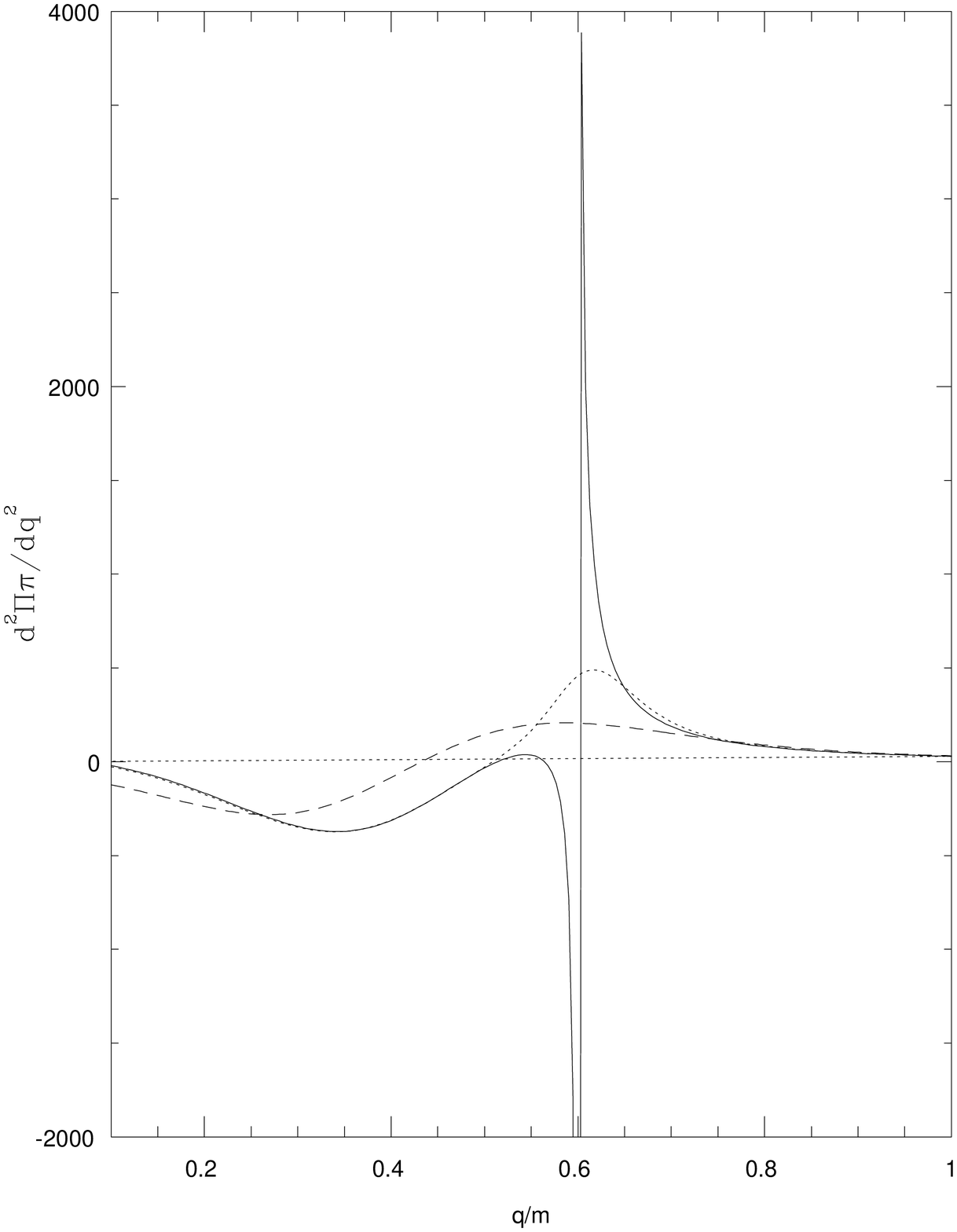}{Second derivative of the pion polarization as a
function of $q/m$ for three temperatures at saturation density : at zero
temperature (solid curve) the Kohn singularity at $q=2 p_f$ is apparent. As
temperature increases, the curve becomes smooth. We showed this behavior
for $T=5$ MeV (short dashes) and  $T=20$ MeV (long dashes).}{Fig. 2}

\mafigura{6cm}{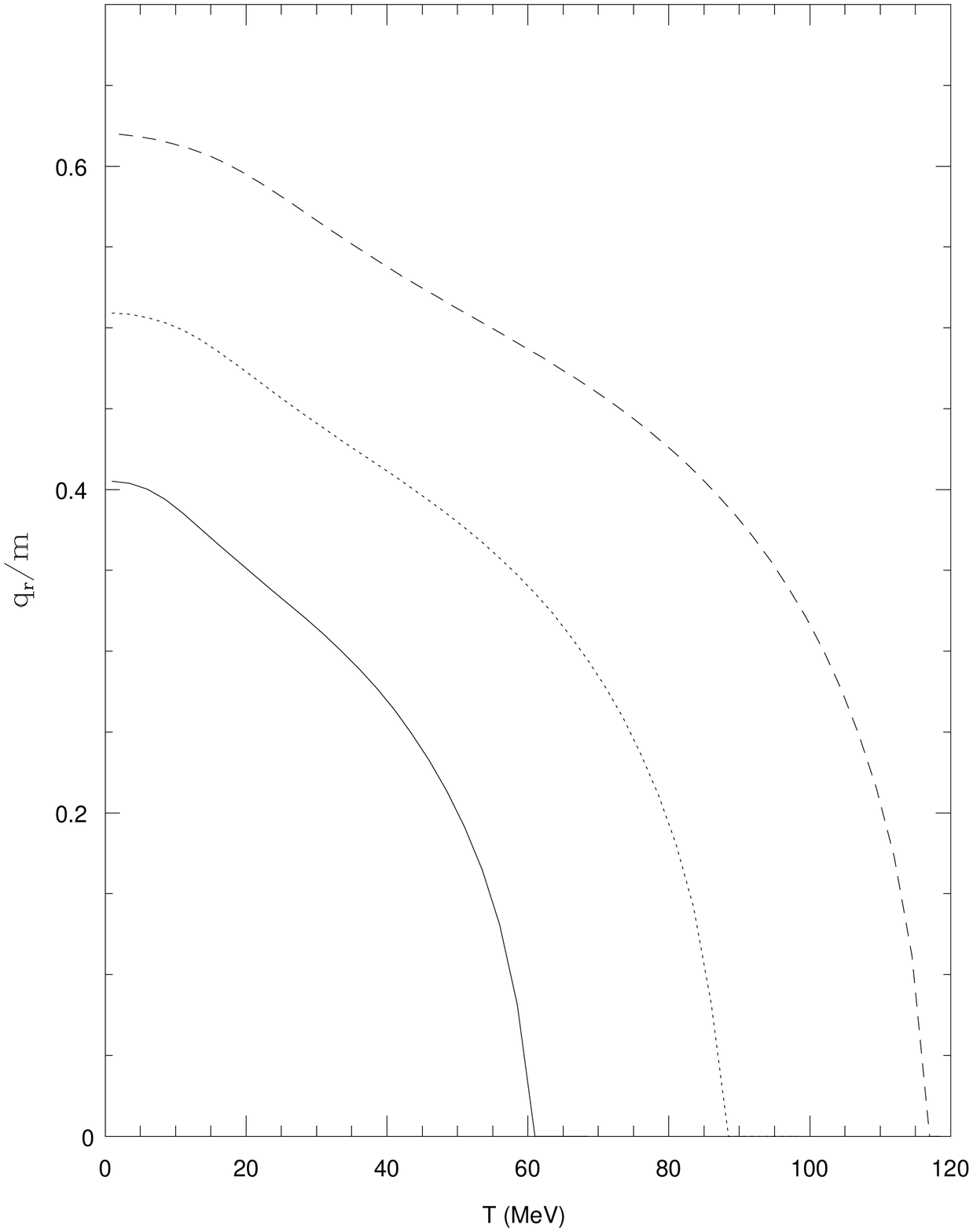}{Evolution of the real part of the Yukawa pole, for
three values of the density, as temperature increases. The curves
correspond to $p_f=0.3$ (solid curve), $p_f=0.35$ (dotted) and $p_f=0.4$
(dashed), which are values of the nucleon Fermi momentum at $T=0$.}{Fig. 3}

\mafigura{6cm}{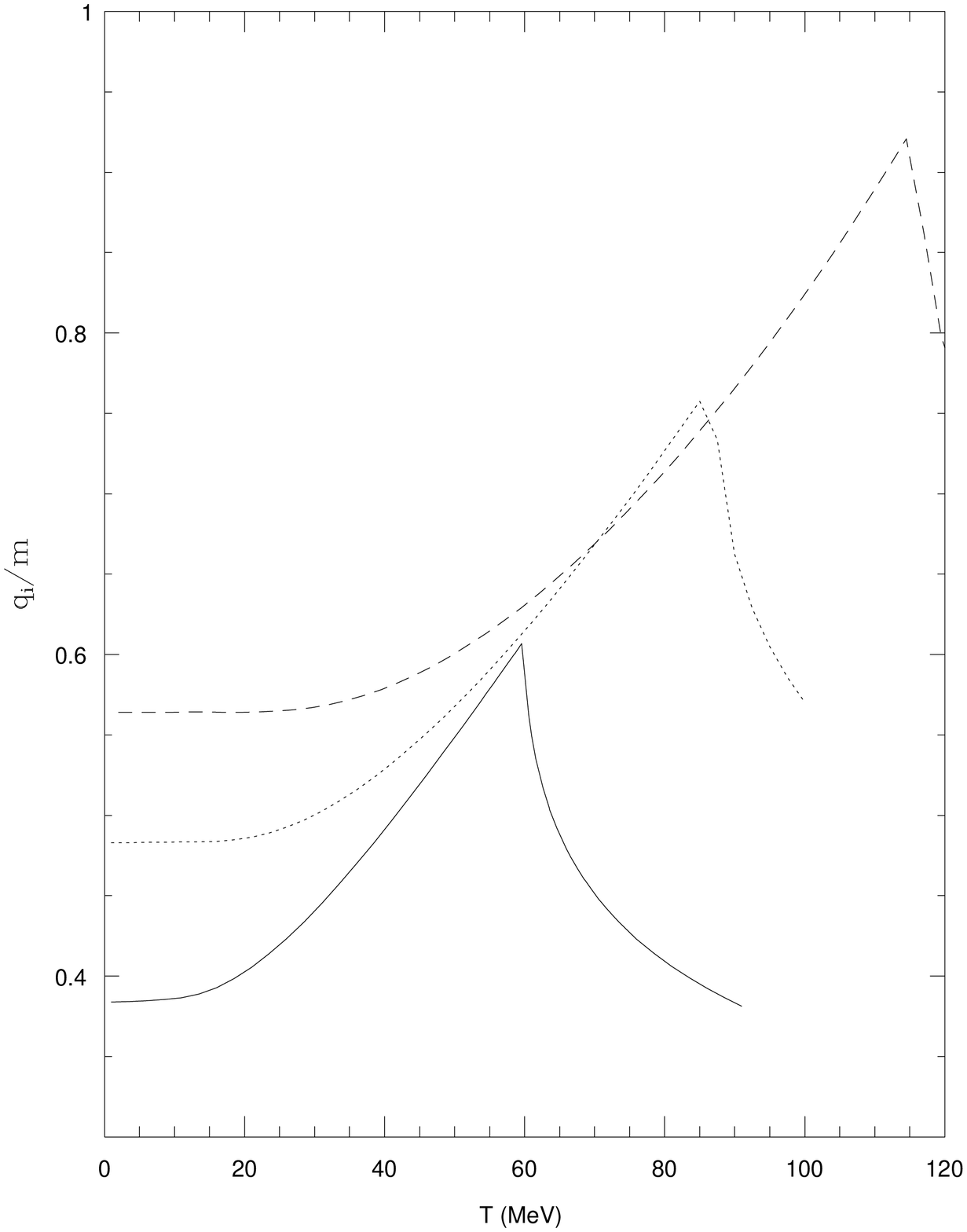}{Same as before, for the imaginary part of the pole.}
{Fig. 4}

\mafigura{6cm}{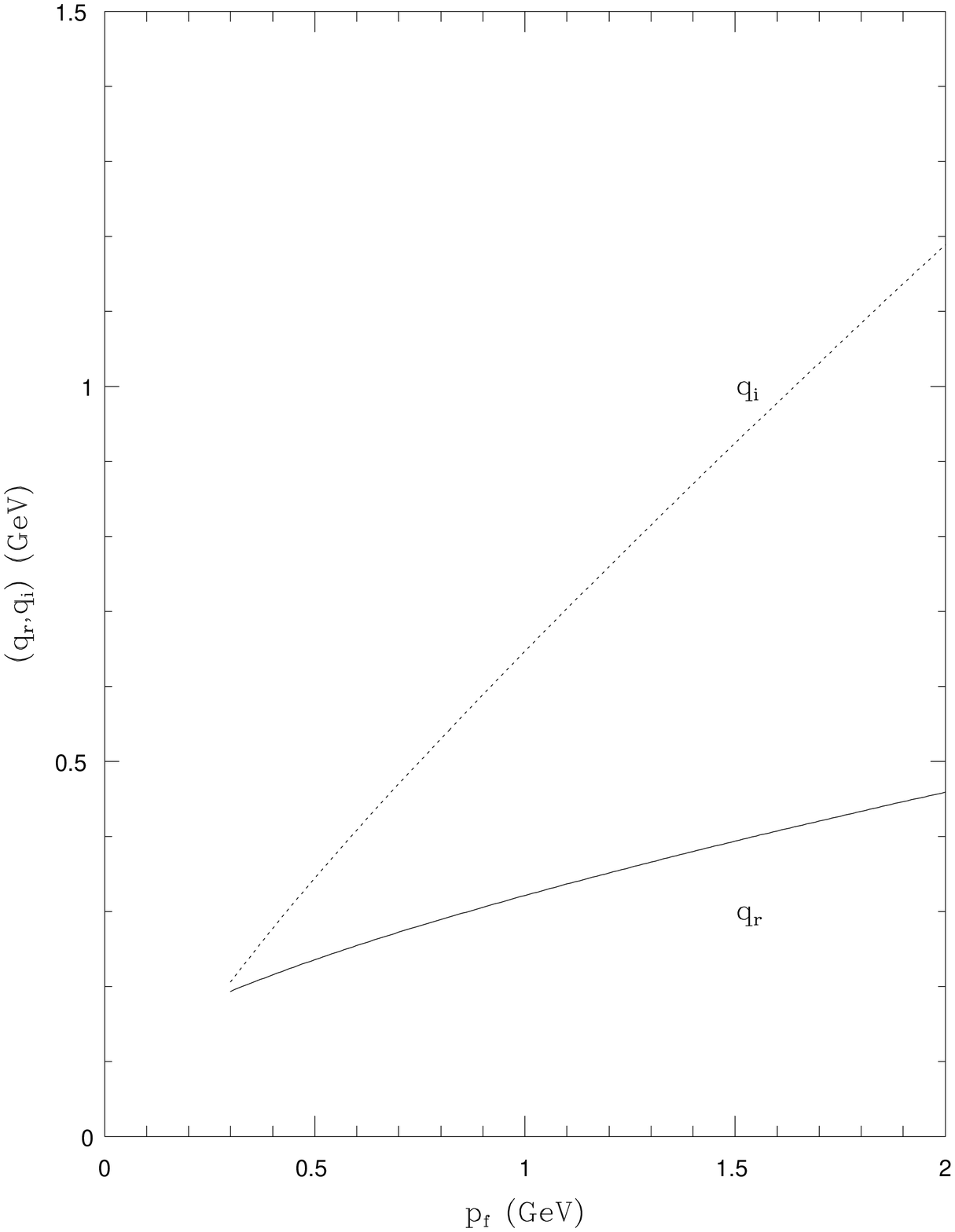}{The real and imaginary parts of the Yukawa pole in
the gluon propagator at zero temperature, as a function of the quark Fermi
momentum.}{Fig. 5}

\mafigura{6cm}{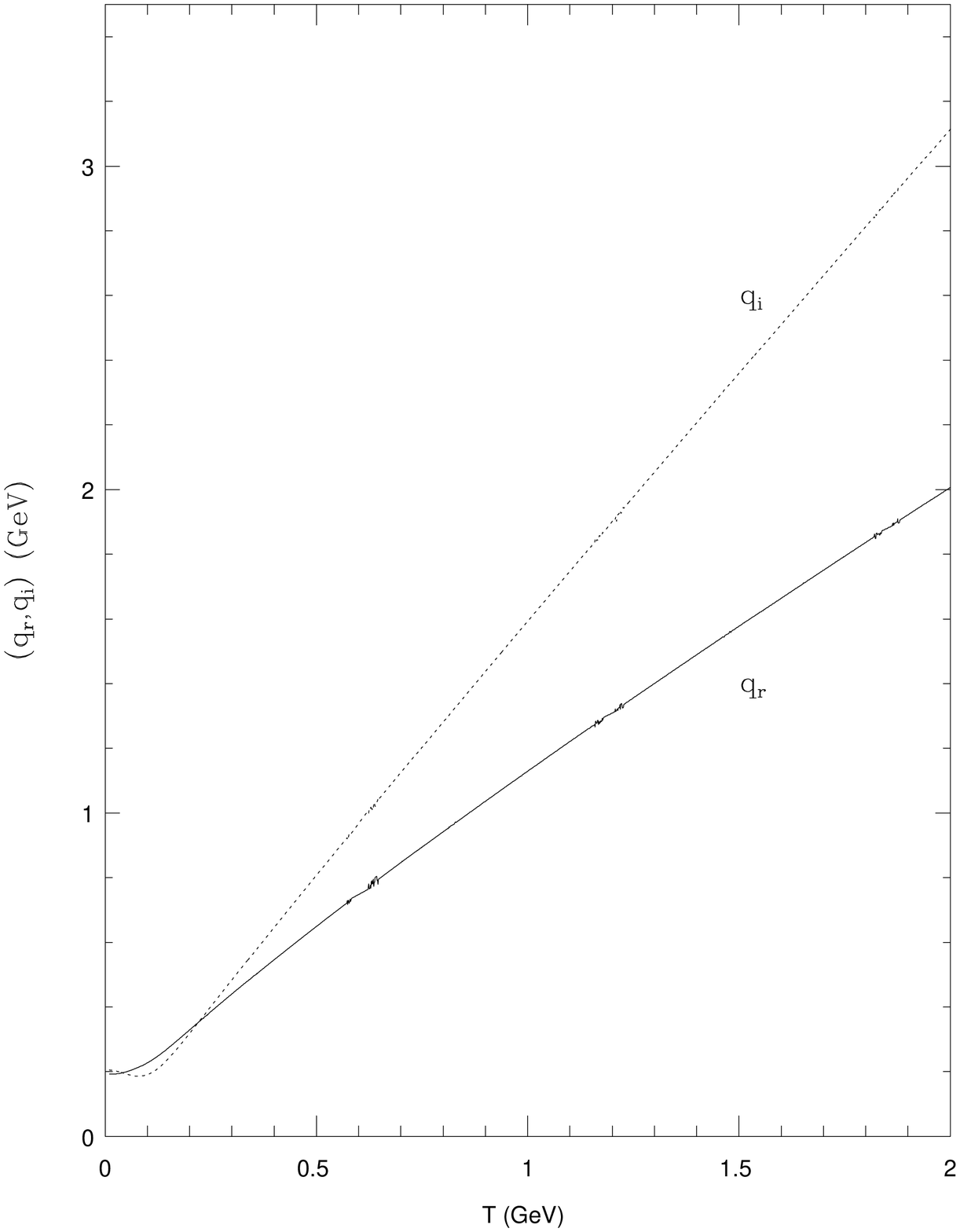}{Evolution of the pole of the gluon propagator as
temperature changes, for a density corresponding to $p_f=300$ MeV at $T=0$.}
{Fig. 6}

\mafigura{6cm}{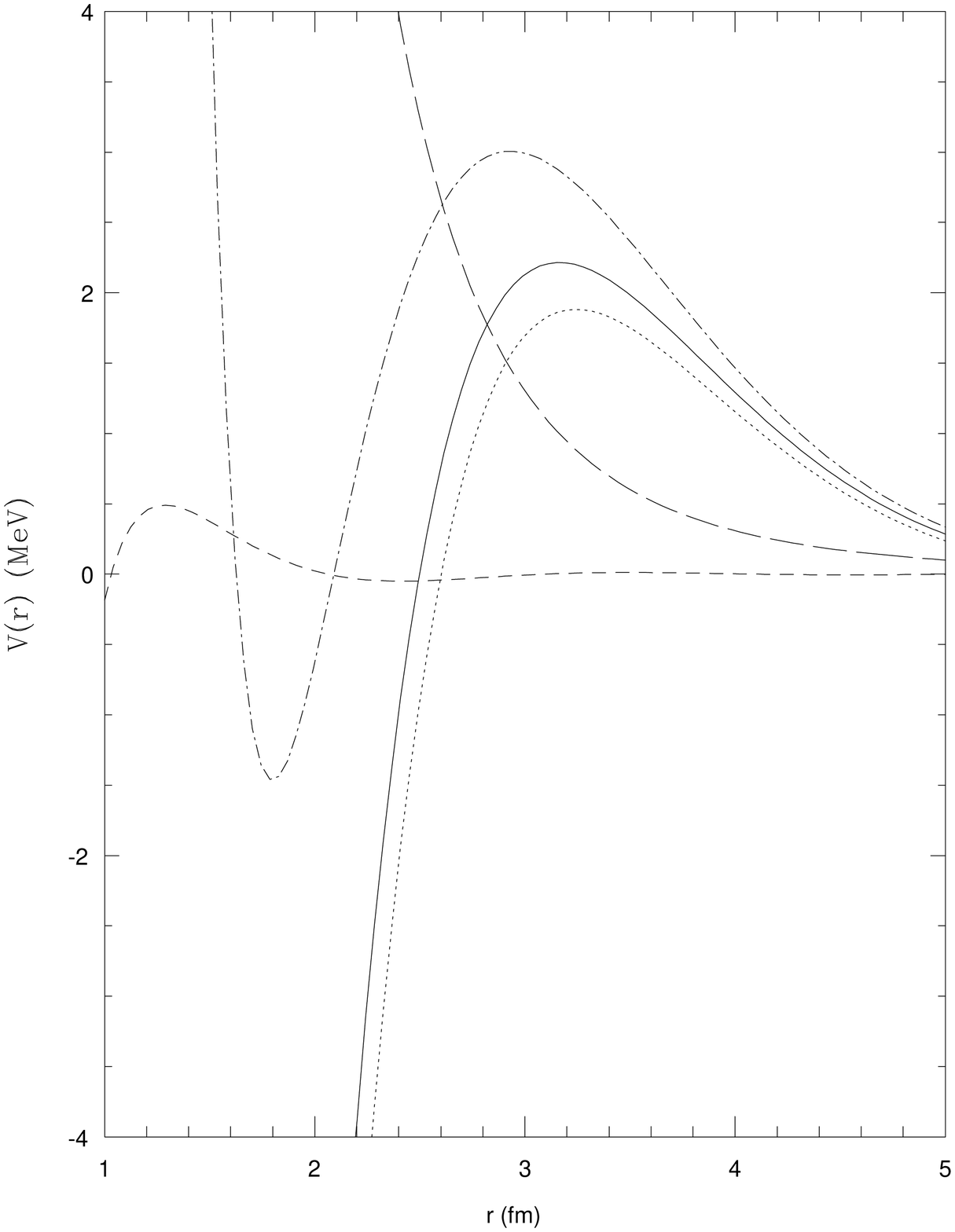}{Different components of the potential $V(r)$ in a
QCD plasma, at zero temperature and $p_f=300$ MeV. Solid line : result of the
numerical integration of Eq. (\ref{vqcd}). The dotted line shows the Yukawa
component, as given by Eq. (\ref{vyuk}) . Also shown is the contribution from
Eq. (\ref{contribim}) (long dashes) and the Friedel component (short dashes). 
Finally, the dotted-dashed line is
obtained by the sum of the Friedel, Yukawa and Eq. (20).}{Fig. 7} 

\mafigura{6cm}{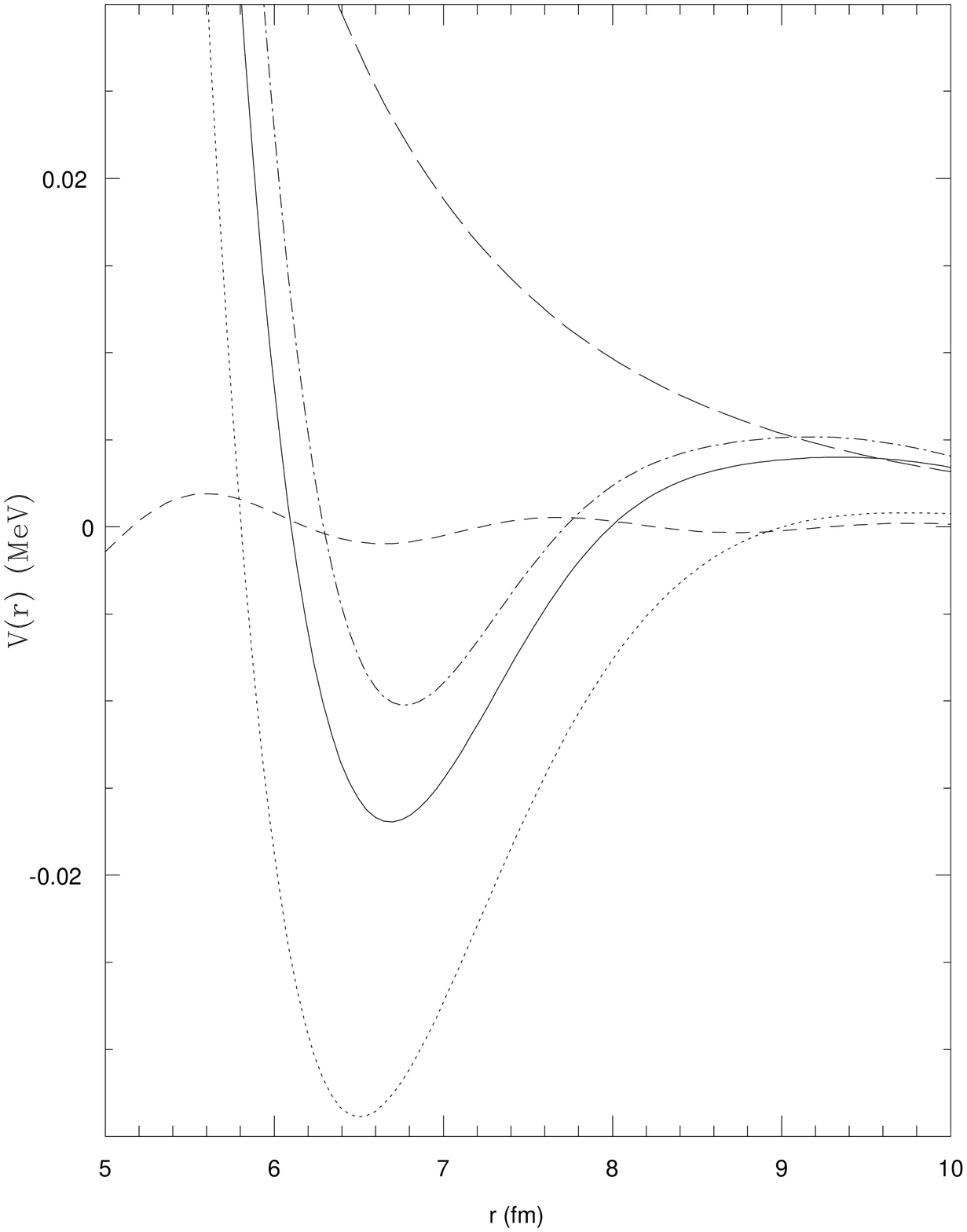}{Same as Fig. 7, for $r$ between 5 and 10 fm.}{Fig. 8}

\mafigura{6cm}{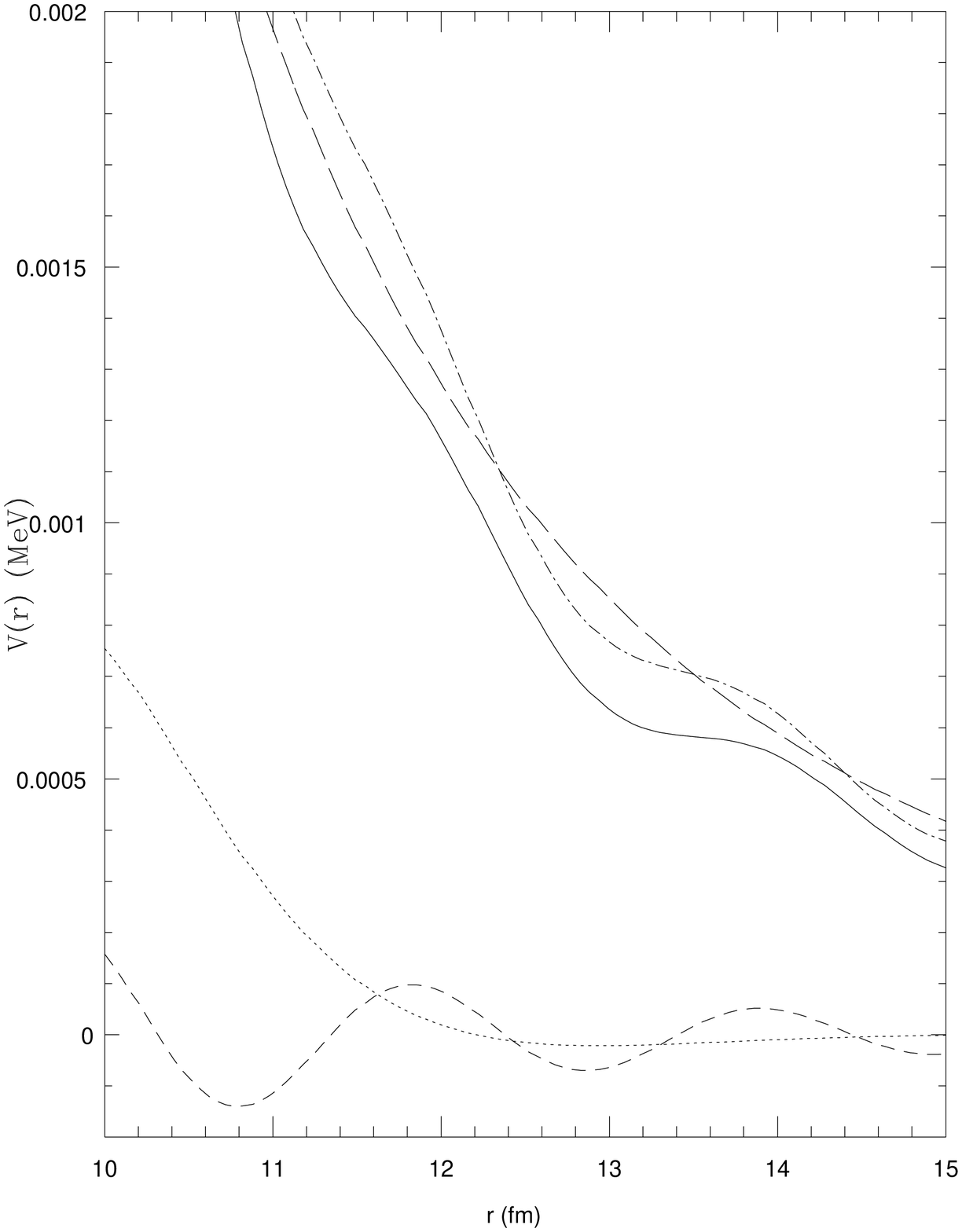}{Same as Fig. 7, for $r$ between 10 and 15 fm.}{Fig. 9}

\end{document}